\documentclass{emulateapj}
\usepackage{natbib}

\newcommand{\teff}{T$_{\rm eff}$}

\newcommand{\ergs}{erg~s$^{-1}$}

\newcommand{\ujy}{$\mu$Jy}

\newcommand{\name}{2MASS~J13153094$-$2649513}
\newcommand{\namesh}{2MASS~J1315$-$2649}

\slugcomment{Submitted to ApJ Letters 27 August 2012; Accepted 14 November 2012}

\shorttitle{Radio Emission from 2MASS~J1315$-$2649AB}
\shortauthors{Burgasser et al.}

\begin{document}

\title{Detection of Radio Emission from the Hyperactive L Dwarf 2MASS~J13153094$-$2649513AB}

\author{
Adam J.\ Burgasser\altaffilmark{1},
Carl Melis\altaffilmark{1,2},
B.\ Ashley Zauderer\altaffilmark{3}, and
Edo Berger\altaffilmark{3}
}

\altaffiltext{1}{Center for Astrophysics and Space Science, University of California San Diego, La Jolla, CA 92093, USA; aburgasser@ucsd.edu}
\altaffiltext{2}{Joint NSF AAPF Fellow and CASS Postdoctoral Fellow}
\altaffiltext{3}{Harvard-Smithsonian Center for Astrophysics, 60 Garden Street, Cambridge, MA 02138, USA}

\begin{abstract}
We report the detection of radio emission from the unusually active L5e + T7 binary 2MASS J13153094$-$2649513AB made with the Australian Telescope Compact Array.  Observations at 5.5~GHz reveal an unresolved source with a continuum flux of 370$\pm$50~{\ujy}, corresponding to a radio luminosity of
$L_{rad} = {\nu}L_{\nu}$ = (9$\pm$3)$\times$10$^{23}$~{\ergs} and $\log_{10}{L_{rad}/L_{bol}}$ = $-$5.44$\pm$0.22. 
No detection is made at 9.0~GHz to a 5$\sigma$ limit of 290~{\ujy}, consistent with a power law spectrum $S_{\nu} \propto \nu^{-\alpha}$ with $\alpha \gtrsim 0.5$.
The emission is quiescent, with no evidence of variability or bursts over 3~hr of observation, and no measurable polarization (V/I $<$ 34\%).
2MASS~J1315$-$2649AB is one of the most radio-luminous ultracool dwarfs detected in quiescent emission to date, 
comparable in strength to other cool sources detected in outburst.
Its detection indicates no decline in radio flux through the mid-L dwarfs. 
It is unique among L dwarfs in having
strong and persistent H$\alpha$ and radio emission, indicating
the coexistence of a cool, neutral photosphere (low electron density) and a highly active chromosphere (high electron density and active heating). 
These traits, coupled with the system's mature age and substellar secondary, makes
2MASS~J1315$-$2649AB an important test for proposed radio emission mechanisms 
in ultracool dwarfs.
\end{abstract}

\keywords{
stars: chromospheres ---
stars: individual (\objectname{2MASS~J13153094$-$2649513}) --- 
stars: brown dwarfs ---
stars: low mass ---
stars: magnetic field
}

\section{Introduction}

The origin of radio emission from ultracool dwarfs (UCDs; late M-, L- and T-type stars and brown dwarfs with T$_{eff} \lesssim$ 3000K) remains one of the great mysteries in our understanding of these cool, low-mass objects.  The first radio detection of the M9 brown dwarf LP~944-20 \citep{2001Natur.410..338B} was unexpected given the precipitous decline in the strength and incidence of optical
and X-ray emission among late-M and L dwarfs 
($\log{L_{H\alpha}/L_{bol}}$ $\approx$ $\log{L_{X}/L_{bol}}$ $\approx -4$ to $\lesssim -6$;
\citealt{2000AJ....120.1085G,2006A&A...448..293S,2007AJ....133.2258S}) and the correlation between X-ray and radio flux among many stellar sources (i.e., the Guedel-Benz relation, $\log_{10}{L_{\nu,rad}/L_X} \approx -15.5$~Hz$^{-1}$; \citealt{1993ApJ...405L..63G,1994A&A...285..621B}).
In the chromospheric evaporation model of solar and stellar flares \citep{1980ApJ...242..336M,2006ApJ...644..484A}, 
the radio/X-ray correlation is attributed to the heating and evaporation of chromospheric plasmas (X-ray emission) by accelerated electrons (radio emission) via the Neupert effect \citep{1968ApJ...153L..59N}.
The dozen UCDs detected in the radio to date \citep{2002ApJ...572..503B,2006ApJ...648..629B,2005ApJ...626..486B,2007ApJ...658..553P,2008A&A...487..317A,2011ApJ...741...27M,2012ApJ...746...23M,2012ApJ...747L..22R} violate the
Guedel-Benz relation by orders of magnitude, suggesting a breakdown in this mechanism.
UCD radio emission also exhibits a broad range of behaviors,
including low-polarization quiescent emission (e.g., \citealt{2002ApJ...572..503B}),
sporadic variability (e.g., \citealt{2007A&A...472..257A}),
periodic variability synched with rotation (e.g., \citealt{2005ApJ...627..960B,2011ApJ...741...27M}), 
highly polarized bursts (e.g., \citealt{2005ApJ...626..486B}), 
and ``pulsar-like'' rotationally-synched coherent emission (e.g., \citealt{2007ApJ...663L..25H,2009ApJ...695..310B}).  
The complexity of these behaviors, and the decoupling of radio, X-ray and H$\alpha$ emission trends, has 
stimulated new theoretical work on the origin of magnetic emission in very cool stars, brown dwarfs and exoplanets
(e.g., \citealt{2009ApJ...699L.148S,2010A&A...522A..13R,2012ApJ...746...99K}).

While the decline in optical emission in UCDs is a general trend, a small number  of unusually ``hyperactive'' cool dwarfs have been identified, whose strong and persistent H$\alpha$ emission also remains a mystery
\citep{2000AJ....120..473B,2003AJ....125..343L,2007AJ....133.2258S}.
One such source is the L5e {\name} (hereafter {\namesh}; \citealt{2002ApJ...564L..89H, 2002ApJ...575..484G}), a high proper motion, very cool dwarf that has exhibited pronounced and sustained H$\alpha$ emission at the level of $\log_{10}{L_{H\alpha}/L_{bol}}$ $\approx$ $-$4 for over a decade \citep{2002ApJ...564L..89H, 2002ApJ...575..484G, 2005A&A...439.1137F,2011ApJ...739...49B}.
The optical emission, which includes Ca~II and alkali resonance lines, is $\sim$100 times stronger than equivalently classified L dwarfs. 
{\namesh} also harbors a resolved T dwarf companion, which appears to be  
too widely separated to induce magnetic interaction \citep{2011ApJ...739...49B}.
Kinematics and coevality analyses indicate that this is a mature system 
(few Gyr) and that {\namesh}A is likely a low mass star just above the hydrogen burning mass limit (M $\sim$ 0.075~M$_{\odot}$).  Its strong optical emission therefore contradicts stellar age-activity trends 
\citep{1995ApJ...450..401F},
again pointing to new magnetic behavior in the UCD regime.

In this Letter, we report the detection of quiescent, unpolarized radio emission from {\namesh} at 5.5 GHz based on observations from the Australian Telescope Compact Array (ATCA).  The radio luminosity of this source exceeds all other UCDs detected in quiescent emission to date, but is consistent with a lower radio to H$\alpha$ luminosity fraction than other L dwarfs.
In Section~2 we summarize the observations, data analysis and resulting measurements.
In Section~3 we use these data to infer the properties of the emitting region and compare to other UCD emitters.
In Section~4 we discuss the possible origins of this emission and propose future observations.

\section{Observations}

{\namesh} was observed with ATCA in the compact hybrid H214 configuration (baselines of 0.082--5.94 km) on 14 July 2011 (UT). Continuum observations were taken in dual-sideband mode simultaneously at 5.5 GHz (C-band) and 9.0 GHz (X-band). The Compact Array Broadband Backend (CABB; \citealt{2011MNRAS.416..832W}) was used, providing 2\,GHz bandwidth per observing frequency in 2048 channels of 1\,MHz width each.
The source was tracked for a total of 3~hr in 5~min and 10~min intervals, interspersed with observations of the gain calibrator QSO\,B1255-316.
The quasar QSO\,B1934-638 was used
for primary flux calibration, while QSO\,B0823-500 was used for bandpass calibration.
Data were reduced using the Astronomical Image Processing System package (AIPS; \citealt{2003ASSL..285..109G}) following best practices for wide-band data reduction.

\begin{figure*}
\epsscale{1}
\plottwo{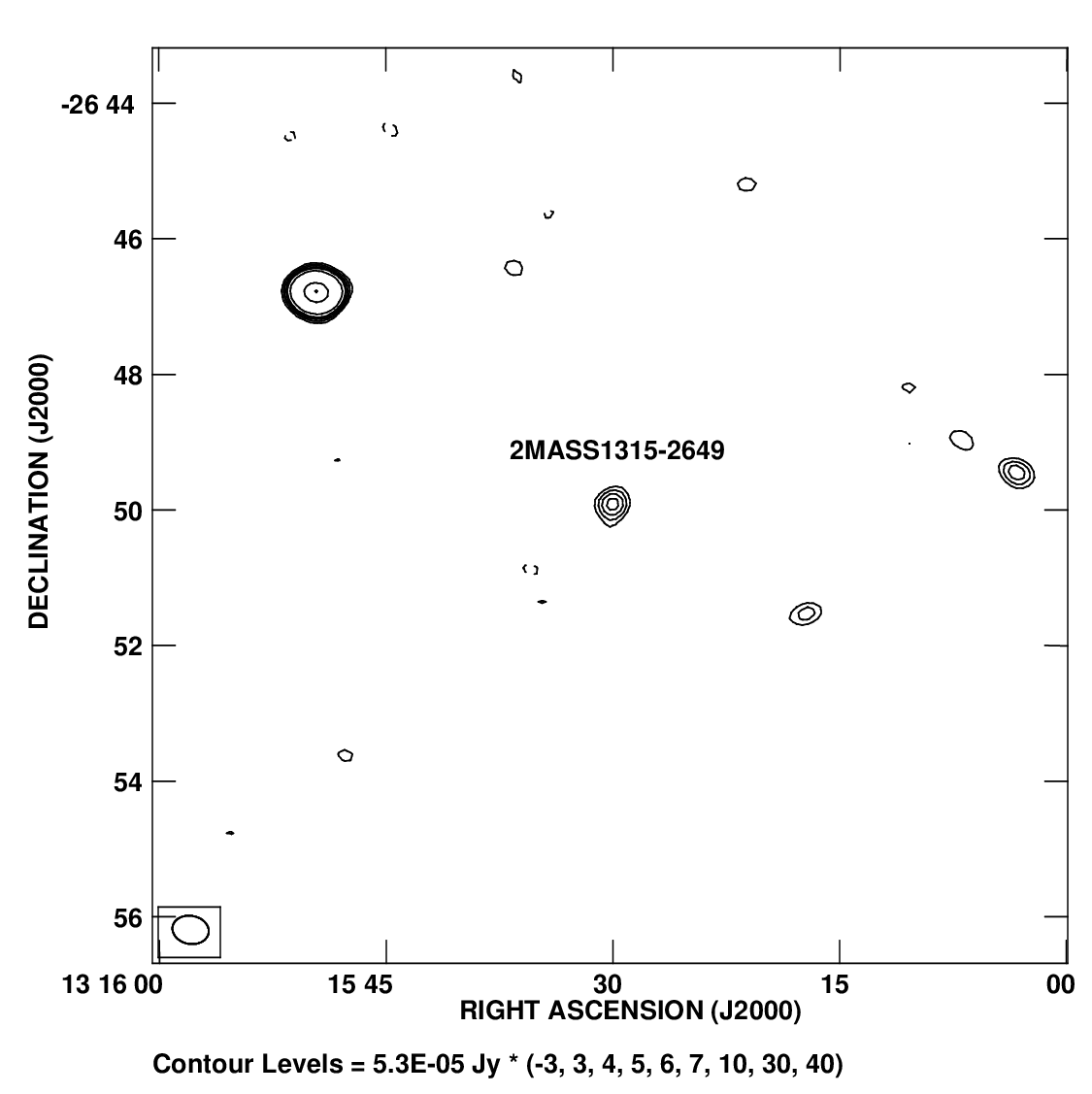}{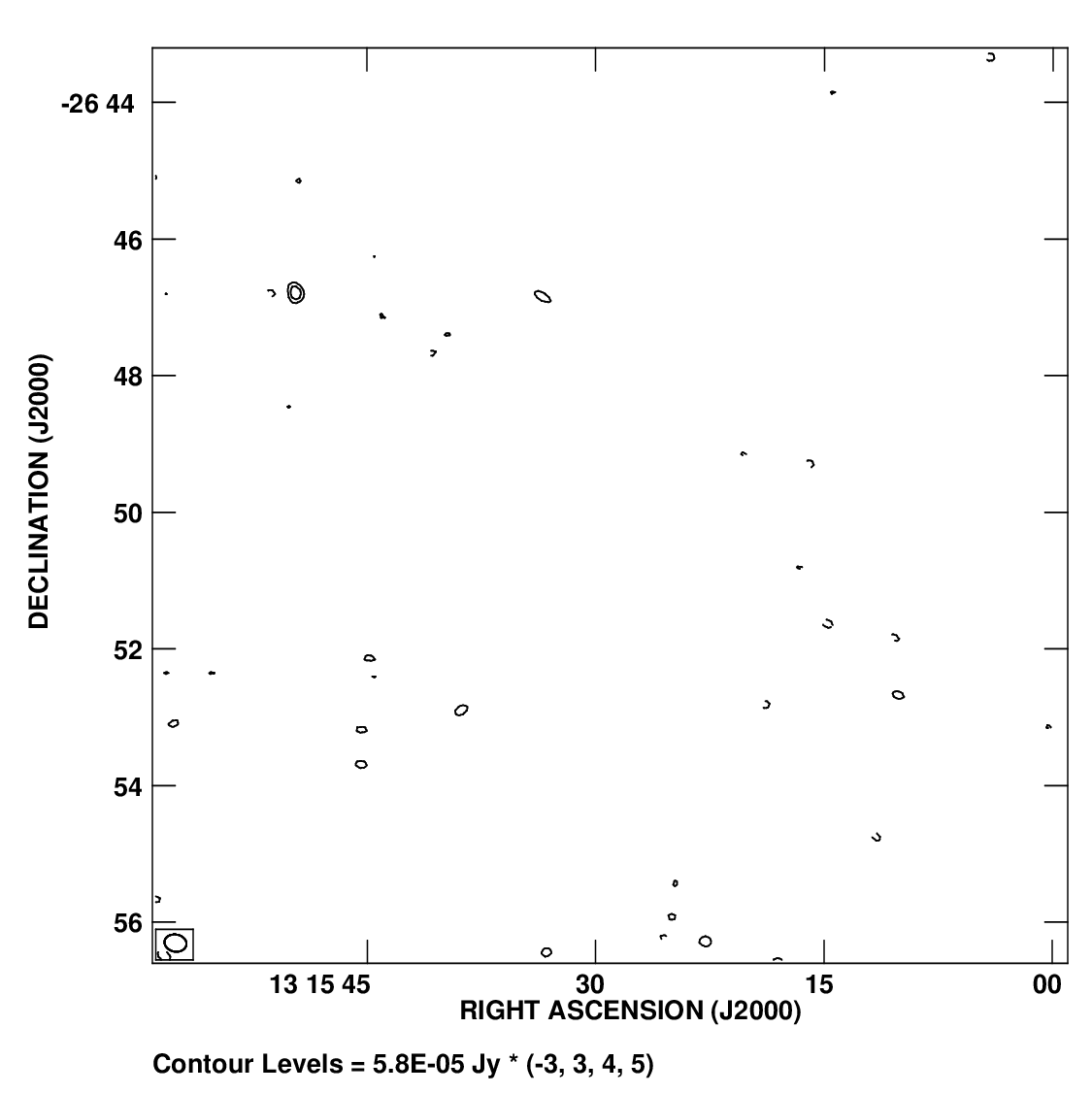}
\caption{13$\arcmin\times$13$\arcmin$ contour images of the {\namesh} field in integrated 5.5~GHz (left) and 9.0~GHz (right) flux density, based on our ATCA observations.  Contours are spaced at listed intervals of the noise (53~{\ujy}~beam$^{-1}$ in the 5.5~GHz data, 58~{\ujy}~beam$^{-1}$ in the 9.0~GHz data; negative contours are dashed) and the beam shape is indicated in the lower left corners. {\namesh} is detected at 5.5~GHz with 7.4$\sigma$ confidence.}
\label{fig:image}
\end{figure*}

Images of the {\namesh} field from the integrated broadband 5.5~GHz and 9.0~GHz data are shown in Figure~\ref{fig:image},
with synthesized beam sizes ($B_{\nu}$) of $\approx$30$\arcsec$ and $\approx$25$\arcsec$, respectively.
Several significant sources are seen in each field, but in the 5.5~GHz data there is one source centered at\footnote{Equinox J2000 coordinates}
(13$^h$15$^m$29$\fs$99, $-$26$\degr$49$\arcmin$55$\farcs$7)$\pm$(0$\fs$12, 2$\farcs$0) that is coincident with the proper-motion-corrected position of {\namesh} to within 2$\sigma$.\footnote{The source located 5$\farcm$4 northeast of {\namesh} is  NVSS~J131549-264647 (\citealt{1998AJ....115.1693C}; $S_{1.4}$ = 13.8$\pm$0.6~mJy, $S_{5.5} \approx 3$~mJy).  Its dirty beam pattern was cleaned using standard AIPS routines, and does not contribute to the flux detected at the position of {\namesh}.} 
 This source has an integrated flux density of $S_{5.5}$ = 370$\pm$50~{\ujy}, a 7.4$\sigma$ detection.
We rule out source confusion as the origin of this emission by noting that the deep 5~GHz survey  of the Lockman Hole by \citet{2003A&A...398..901C} measured a source density of
$N$ = 0.04~arcmin$^{-2}$ for $S_{5.0} >$ 350~{\ujy}, implying a confusion probability $1-e^{-NB_{5.5}^2}$ $\approx$ 1\% for this source.
No equivalent source is detected in the 9.0~GHz data to a 5$\sigma$ limiting flux of 290~{\ujy}.

The 5.5~GHz emission from {\namesh} appears to be roughly constant over the observing period. 
Dividing the data into two equally-spaced periods yields consistent fluxes of 430$\pm$60~{\ujy} and 340$\pm$70~{\ujy}.  Further division fails to yield a significant detection on timescales of 10~s to 1~hr, indicating a bursting flux limit of $<$0.5~mJy ($<$1.3~mJy) for a 30~min (5~min) event, timescales typical of previously observed radio bursts (e.g., \citealt{2002ApJ...572..503B,2005ApJ...626..486B,2007ApJ...663L..25H,2009ApJ...695..310B}). 
There is a suggestion of spectral structure in the emission, as division of the time-integrated flux is slightly dominated by the lowest frequency channels ($\lesssim$5.2~GHz), consistent with a negative power-law spectral slope (see below).  However, these differences are not statistically significant.  
We also detect no significant polarization in the emission, with a Stokes $V/I$ 3$\sigma$ limit of $\lesssim$34\%.

\section{Characterizing the Radio Emission of {\namesh}}

The magnitude of the radio emission from {\namesh} is remarkable, particularly given the late spectral types of its components. For the following, we assume the emission arises from the L5 primary (see Section~4 for discussion on possible contribution from the secondary).  The absolute radio intensity at 5.5~GHz is
$L_{\nu,rad}$ = $4{\pi}d^2S_{5.5}$ = (1.6$\pm$0.6)$\times$10$^{14}$ erg~s$^{-1}$~Hz$^{-1}$
based on the estimated $d$ = 19$\pm$3~pc distance of the {\namesh} system.
Parameterizing the radio power as $L_{rad} = {\nu}L_{\nu}$, we find\footnote{Alternately, adopting a spectral flux distribution that peaks around 5.5~GHz, with $S_{\nu} \propto \nu^{2.5}$ for $\nu <$ = 5.5~GHz and $S_{\nu} \propto \nu^{-1.5}$ for $\nu >$ 5.5 GHz \citep{1985ARA&A..23..169D}, and integrating over 6 decibels about 5.5 GHz, we derive a statistically equivalent $L_{rad}$ = (1.5$\pm$0.5)$\times$10$^{24}$~erg~s$^{-1}$.} 
$L_{rad}$ = (9$\pm$3)$\times$10$^{23}$~erg~s$^{-1}$, corresponding to 
$\log_{10}{L_{rad}/L_{bol}}$ = $-$5.44$\pm$0.22 assuming a bolometric luminosity of $\log_{10}{L_{bol}/L_{\sun}}$ = $-$4.19$\pm$0.16 for the primary \citep{2011ApJ...739...49B}.
The lack of emission at 9.0~GHz implies a power-law slope in the radio flux $S_{\nu} \propto \nu^{-\alpha}$ with $\alpha \gtrsim 0.5$. This index is broadly consistent with 
the radio spectra of other UCDs \citep{2009ApJ...700.1750O,2011ApJ...735L...2R}, with the exception of the flat-spectrum radio source 2MASS~J1314+1320 \citep{2011ApJ...741...27M}.

\begin{deluxetable}{lc}
\tabletypesize{\small}
\tablecaption{Radio Properties of {\name} \label{tab:radio}}
\tablewidth{0pt}
\tablehead{
\colhead{Parameter} &
\colhead{Value} \\
}
\startdata
Right Ascension\tablenotemark{a}  & 13$^h$15$^m$29$\fs$99$\pm$0$\fs$12 \\
Declination\tablenotemark{a}  & $-$26$\degr$49$\arcmin$55$\farcs$7$\pm$2$\farcs$0 \\
$S_{5.5}$ ({\ujy}) & 370$\pm$50 \\
$S_{9.0}$ ({\ujy}) & $<$290\tablenotemark{b} \\
$\alpha$\tablenotemark{c} & $\gtrsim$0.5 \\
$\log_{10}{L_{5.5}}$ (erg~s$^{-1}$~Hz$^{-1}$) & 14.18$\pm$0.15 \\
$\log_{10}{{\nu}L_{\nu}}$ (erg~s$^{-1}$) & 23.92$\pm$0.15 \\
$\log_{10}{{\nu}L_{\nu}/L_{bol}}$ & -5.44$\pm$0.22 \\
$\log_{10}{{\nu}L_{\nu}/L_{H\alpha}}$ & -1.3$\pm$0.5\tablenotemark{d} \\
\enddata
\tablenotetext{a}{Equinox J2000 coordinates on Julian data 2455757.}
\tablenotetext{b}{5$\sigma$ upper limit.}
\tablenotetext{c}{Assuming $S_{\nu} \propto \nu^{-\alpha}$.}
\tablenotetext{d}{Accounting for $\pm$0.4~dex variation in H$\alpha$ EW measurements reported in the literature.}
\end{deluxetable}

\begin{figure*}
\epsscale{1.1}
\plottwo{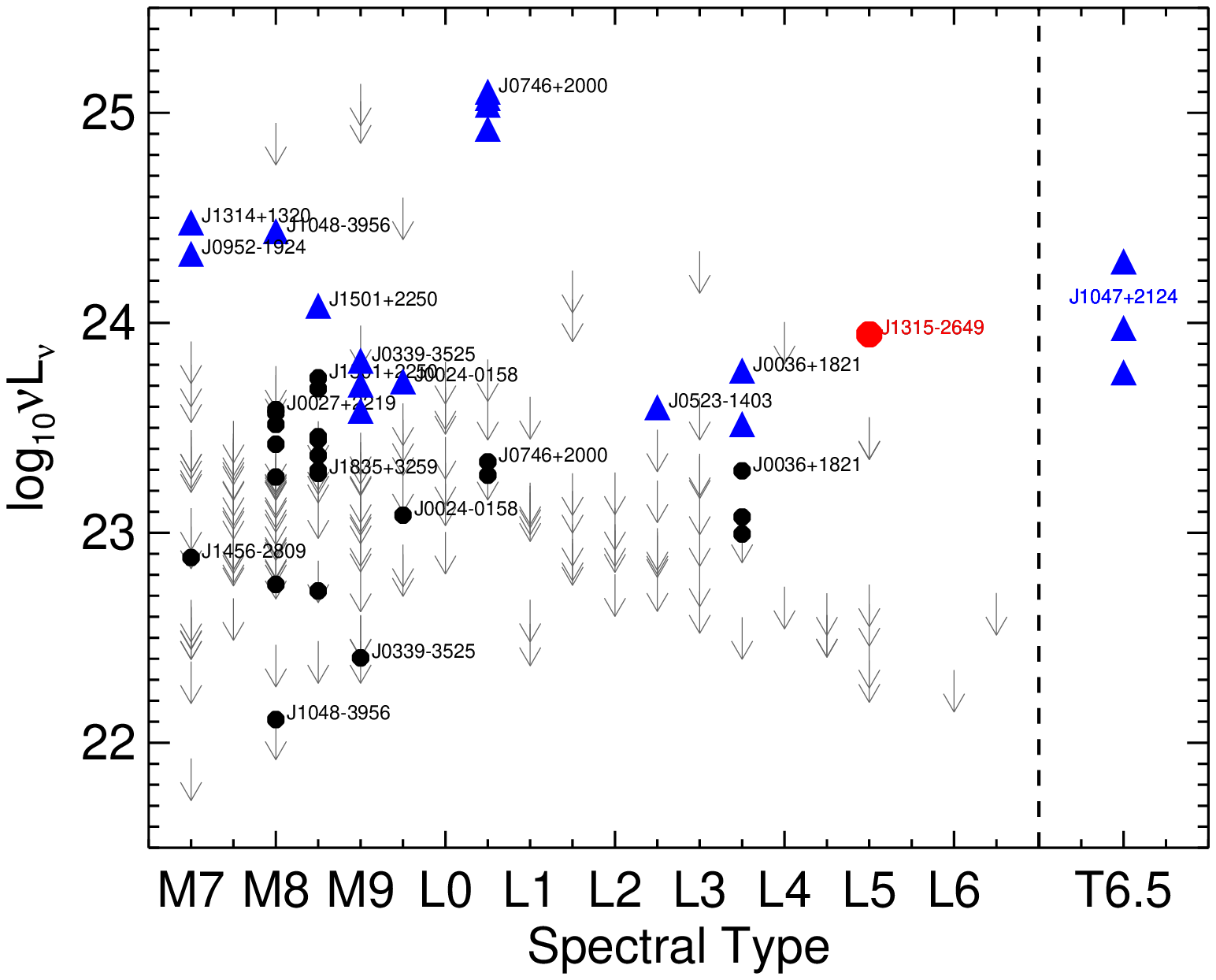}{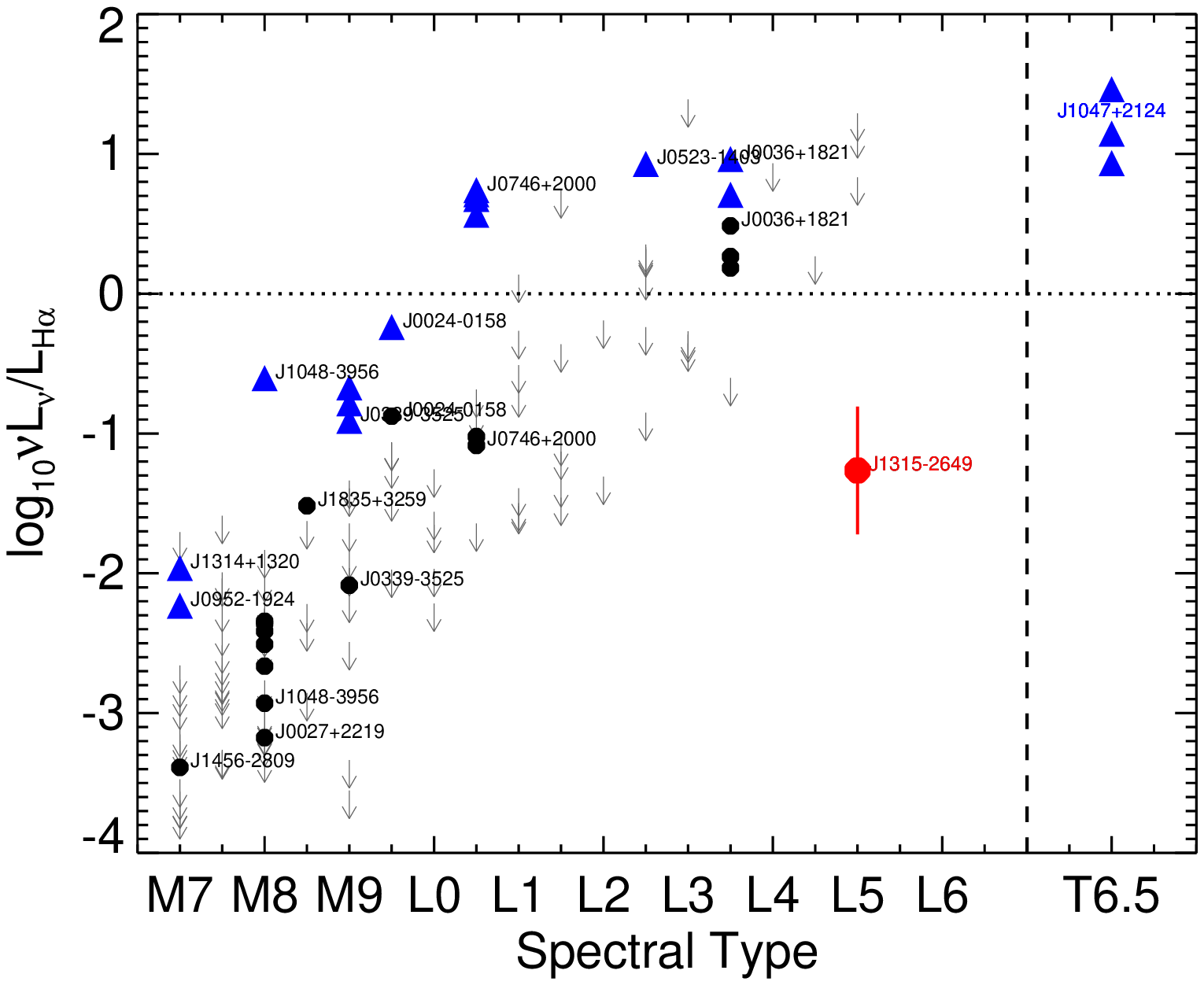}
\caption{{\it (Left):} Radio luminosity (${\nu}L_{\nu}$) versus spectral type for M7 and later sources, based on the compilation of \citet{2012ApJ...746...23M} supplemented with results from \citet{2006ApJ...637..518O,2009ApJ...700.1750O,2012ApJ...747L..22R}; and this study.  
Detections in quiescent (black circles) and peak outburst (blue triangles) are labelled by source sexigessimal coordinates; upper limits are indicated by downward arrows.  Quiescent emission from {\namesh}, with $\log{{\nu}L_{\nu}} = 23.92{\pm}0.15$ at 5.5~GHz, is indicated by the red circle.
Spectral type uncertainties are typically $\pm$0.5 subtypes.
{\it (Right):}  Ratio of radio to H$\alpha$ luminosity as a function of spectral type (same symbols).  Only sources with H$\alpha$ detections are plotted.  
The relative emission of {\namesh}, $\log{{\nu}L_{\nu}/L_{H\alpha}} = -1.3{\pm}0.4$ (uncertinaty includes scatter in the observed H$\alpha$ measurements), is distinct from the other L dwarfs and more consistent with a late M dwarf. 
}
\label{fig:ratios}
\end{figure*}

To place this emission in context, we compare the radio luminosity of {\namesh} to other UCDs in Figure~\ref{fig:ratios}.  We find {\namesh} to be the most radio-luminous quiescent emitter among this group, comparable to the peak fluxes detected among UCD radio outbursts.  Its brightness is consistent with no appreciable decline in radio power with spectral type well into the L dwarf class, at least for radio-loud UCDs \citep{2007A&A...471L..63A,2010ApJ...709..332B}.
No X-ray measurement has yet been made of {\namesh}, but we can infer that it likely violates the Guedel-Benz relation since it would require an X-ray luminosity $\log_{10}{L_X}$ $\approx$ 29.7~erg~s$^{-1}$, or $\log_{10}{L_X/L_{bol}} > 0$ (this is also true for the secondary).  Active M and L dwarfs typically have $\log_{10}{L_X/L_{H\alpha}} \approx 0-1$ \citep{2010ApJ...709..332B}, so even with its strong optical emission {\namesh} likely has $\log_{10}{L_X/L_{bol}} \approx -3.5$ and $\log_{10}{L_{\nu,rad}/L_X} \approx -11.8$~Hz$^{-1}$, violating the Guedel-Benz relation by nearly four orders of magnitude.

Figure~\ref{fig:ratios} also displays the (asynchronous) ratio of radio and H$\alpha$ luminosities for UCDs observed to date.  This ratio steadily increases from late-M to mid-L types, in both quiescent and flaring fluxes, reflecting the decline in optical emission and plateau in radio emission with later spectral type.  Nearly all L and T dwarfs appear to emit more nonthermal power in the radio than in optical lines.  {\namesh} is a obvious exception; its strong H$\alpha$ emission implies $\log_{10}{\nu{L_{\nu}}/L_{H\alpha}} = -1.3{\pm}0.5$ (taking into account scatter in the H$\alpha$ measurements), a ratio more in line with mid- and late-M dwarfs.  Given that no later-type source has yet been detected in quiescent flux (the T6.5 2MASS~J1047+2124 has only been detected in outburst; \citealt{2012ApJ...747L..22R}), we can only speculate as to whether this measurement reflects a change in the nonthermal emission spectrum of UCDs at lower temperatures or the unique nature of {\namesh} itself.

\section{Origins for the Emission}

The lack of measurable polarization or variability during our 3~hr observation of {\namesh}
suggests that its radio emission may be driven by incoherent gyrosychrotron emission
\citep{1985ARA&A..23..169D}, as has been previously proposed for quiescent UCD radio sources 
\citep{2002ApJ...572..503B,2006ApJ...637..518O,2011ApJ...735L...2R}.
This emission peaks at high harmonics of the electron cyclotron frequency,
$\nu_{peak} = s\nu_{c} = 2.8sB$~MHz, where $s$ = 10--100, $B$ is local magnetic field strength in Gauss, and $\nu_c = eB/2{\pi}m_ec$ is the electron cyclotron frequency.   Assuming $\nu_{peak} \lesssim$ 5~GHz implies $B \lesssim 20-200$~G, on par with prior estimates for UCD radio emission regions \citep{2006ApJ...648..629B}.  
Razin-Tsytovich suppression 
also provides a constraint on the electron number density, requiring 
$\nu_{peak} > \nu_p^2/\nu_c \approx 29n_e/B$~Hz, or $n_e \lesssim 10^9-10^{10}$~cm$^{-3}$,
where $\nu_p = (n_ee^2/{\pi}m_e)^{1/2}$ is the electron plasma frequency. 
The emitting region scalelength ($L$) is constrained by the brightness temperature of the emission,
$T_{br} \approx 10^{10}({L/{\rm R}_{Jup}})^{-2}$~K,
which cannot exceed the effective temperature of the emitting electrons\footnote{This value assumes an electron energy distribution $N(E) \propto E^{-\delta}$ with $\delta$ = 2 (consistent with $S_{\nu} \propto \nu^{-0.5}$), $E > 10$~keV,  $s$ = 10--100, and that the emission is viewed perpendicular to the field; see Eqn.~37 of \citet{1985ARA&A..23..169D}.},
{\teff} $\approx 10^9-10^{10}$~K $\approx$ 0.1--1~MeV. 
This constraint yields $L \gtrsim$ 1--3~R$_{Jup}$, or 1--3 stellar radii, a scale that is comparable to resolved radio sources associated with M dwarf coronae
\citep{1997A&A...317..707A} and orders of magnitude smaller than the {\namesh}AB separation ($\approx$10$^4$ radii).
This scale size is also equivalent to the corotation orbital radius\footnote{$R_C$ $\equiv$ $\left(GMP^2/4{\pi}^2\right)^{1/3}$ $\approx$ $5(P/hr)^{2/3}(M/M_{\odot})^{1/3}$~R$_{Jup}$ is the radius beyond which  gravitational force cannot provide the necessary centripetal acceleration for solid body rotation with period $P$.} for a 0.08~M$_{\sun}$ UCD dwarf with a rotation period of 1.6~hr, suggesting that if {\namesh} is a rapid rotator (its rotation period is currently unknown), coronal stripping may play a role in the emission geometry \citep{2000MNRAS.318.1217J,2008ApJ...676.1307B,2011ApJ...735L...2R}.
Assuming the presence of a dipole field that scales as $B \propto R^{-3}$, these values suggest a surface field
of roughly 1--5~kG, on par with Zeeman broadening measurements of late-M and L dwarf photospheres \citep{1996ApJ...459L..95J,2010ApJ...710..924R}.  
Thus, gyrosynchrotron emission is a viable mechanism for the radio emission of {\namesh}, provided that it has a sufficiently large magnetic field strength and structure.

\citet{2007ApJ...663L..25H,2008ApJ...684..644H} have proposed an alternative mechanism for UCD radio emission, electron cyclotron masers (ECM; \citealt{1979ApJ...230..621W}) arising from plasma cavities near magnetic poles.  ECM operates when $\nu_c >> \nu_p$, and thus favors
environments with low electron densities and large field strengths like those of UCD atmospheres.  Indeed, the increasingly neutral photospheres associated with declining photospheric temperatures has been proposed as an explanation for the decline
in optical and X-ray emission \citep{2002ApJ...571..469M}.  ECM provides a natural explanation for the highly-polarized, rotationally-modulated bursts observed in, e.g., TVLM~513-46546 and LSR~J1835+3529, but \citet{2008ApJ...684..644H} also argue that unpolarized, broadband, quiescent emission may arise if ECM radiation is dispersed in the ambient plasma 
or emitted from several discrete maser sources across the stellar surface.  Dispersion may be particularly relevant in the magnetosphere of {\namesh} as its persistent and strong H$\alpha$ emission indicates a heating source for coronal plasma; dispersion may have also enhanced its radio detectability if orientation effects are important \citep{2008ApJ...684..644H}.  On the other hand, if plasma densities in the source region exceed $n_e \approx 2{\times}10^3B^2 \approx 10^9-10^{10}$~cm$^{-3}$ for a 1--3~kG field, both ECM and gyrosynchrotron emission may be suppressed \citep{2001P&SS...49.1137Z}.

The calculations above have implicitly assumed that the emission arises from the L5 primary, but the resolution of ATCA cannot rule out emission from the closely-separated T7 companion (338~mas; \citealt{2011ApJ...739...49B}), which 
has a similar spectral type as the T6.5 radio source 2MASS~J10475385+2124234 (hereafter 2MASS~J1047+2124; \citealt{2012ApJ...747L..22R}). We assert that significant nonthermal emission from the secondary is unlikely, however, as this would require a very high, sustained radio luminosity ($\log_{10}{L_{rad}/L_{bol}} \approx -$3.8, on par with the peak emission during the rapid bursts from 2MASS~J1047+2124), 
as well as an H$\alpha$ luminosity that would exceed its bolometric luminosity ($\log_{10}{L_{H\alpha}/L_{bol}}$ $\approx$ +1.6).  Nevertheless, resolved VLBI radio imaging of this binary could resolve the source(s) of emission, and potentially enable short-term astrometric orbit measurements to constrain both the component masses and the system age.\footnote{{\namesh}AB has a projected semi-major axis of 6.6~AU and estimated orbital period of $\sim$50~yr, implying a circular orbital velocity of 0.8~AU~yr$^{-1}$.  For a relative astrometric precision of $\sim$100~$\mu$as, VLBI could in principle detect orbital motion on a daily basis.} 

The radio emission from {\namesh} may prove to be a critical test of the gyrosynchrotron and ECM models, given the strength of the quiescent emission, the low photospheric temperature (for either component), and evidence of a significant chromosphere.  More extensive monitoring of this source is needed to assess whether and over what time scale the radio emission may be variable, which would constrain the length scale of the emission and hence $T_{br}$; and also determine whether periodic bursting emission is present, direct evidence of a coherent emission process.  Coincident monitoring of H$\alpha$, radio and broadband emission would also probe the connection between photospheric and chromospheric structure; studies of structural connection for other UCDs have so far proven inconclusive \citep{2008ApJ...676.1307B,2010ApJ...709..332B}.  
This oddball L dwarf may hold the key to understanding magnetic emission in UCDs in general. 

\acknowledgments

The authors would like to thank Duty Astronomer
Jay Blanchard for his assistance with the observations,
and Juergen Ott for assistance in our application for ATCA green time. 
C.M. acknowledges support from the National Science Foundation under award No.\ AST-1003318;
E.B. acknowledges support from the National Science Foundation under award No.\ AST-1008361

Facilities: \facility{ATCA}


\begin{thebibliography}{50}
\expandafter\ifx\csname natexlab\endcsname\relax\def\natexlab#1{#1}\fi

\bibitem[{{Alef} {et~al.}(1997){Alef}, {Benz}, \&
  {Guedel}}]{1997A&A...317..707A}
{Alef}, W., {Benz}, A.~O., \& {Guedel}, M. 1997, \aap, 317, 707

\bibitem[{{Allred} {et~al.}(2006){Allred}, {Hawley}, {Abbett}, \&
  {Carlsson}}]{2006ApJ...644..484A}
{Allred}, J.~C., {Hawley}, S.~L., {Abbett}, W.~P., \& {Carlsson}, M. 2006,
  \apj, 644, 484

\bibitem[{{Antonova} {et~al.}(2008){Antonova}, {Doyle}, {Hallinan}, {Bourke},
  \& {Golden}}]{2008A&A...487..317A}
{Antonova}, A., {Doyle}, J.~G., {Hallinan}, G., {Bourke}, S., \& {Golden}, A.
  2008, \aap, 487, 317

\bibitem[{{Antonova} {et~al.}(2007){Antonova}, {Doyle}, {Hallinan}, {Golden},
  \& {Koen}}]{2007A&A...472..257A}
{Antonova}, A., {Doyle}, J.~G., {Hallinan}, G., {Golden}, A., \& {Koen}, C.
  2007, \aap, 472, 257

\bibitem[{{Audard} {et~al.}(2007){Audard}, {Osten}, {Brown}, {Briggs},
  {G{\"u}del}, {Hodges-Kluck}, \& {Gizis}}]{2007A&A...471L..63A}
{Audard}, M., {Osten}, R.~A., {Brown}, A., {Briggs}, K.~R., {G{\"u}del}, M.,
  {Hodges-Kluck}, E., \& {Gizis}, J.~E. 2007, \aap, 471, L63

\bibitem[{{Benz} \& {Guedel}(1994)}]{1994A&A...285..621B}
{Benz}, A.~O., \& {Guedel}, M. 1994, \aap, 285, 621

\bibitem[{{Berger}(2002)}]{2002ApJ...572..503B}
{Berger}, E. 2002, \apj, 572, 503

\bibitem[{{Berger}(2006)}]{2006ApJ...648..629B}
---. 2006, \apj, 648, 629

\bibitem[{{Berger} {et~al.}(2001){Berger}, {Ball}, {Becker}, {Clarke}, {Frail},
  {Fukuda}, {Hoffman}, {Mellon}, {Momjian}, {Murphy}, {Teng}, {Woodruff},
  {Zauderer}, \& {Zavala}}]{2001Natur.410..338B}
{Berger}, E., {Ball}, S., {Becker}, K.~M., {Clarke}, M., {Frail}, D.~A.,
  {Fukuda}, T.~A., {Hoffman}, I.~M., {Mellon}, R., {Momjian}, E., {Murphy},
  N.~W., {Teng}, S.~H., {Woodruff}, T., {Zauderer}, B.~A., \& {Zavala}, R.~T.
  2001, \nat, 410, 338

\bibitem[{{Berger} {et~al.}(2010){Berger}, {Basri}, {Fleming}, {Giampapa},
  {Gizis}, {Liebert}, {Mart{\'{\i}}n}, {Phan-Bao}, \&
  {Rutledge}}]{2010ApJ...709..332B}
{Berger}, E., {Basri}, G., {Fleming}, T.~A., {Giampapa}, M.~S., {Gizis}, J.~E.,
  {Liebert}, J., {Mart{\'{\i}}n}, E., {Phan-Bao}, N., \& {Rutledge}, R.~E.
  2010, \apj, 709, 332

\bibitem[{{Berger} {et~al.}(2008){Berger}, {Basri}, {Gizis}, {Giampapa},
  {Rutledge}, {Liebert}, {Mart{\'{\i}}n}, {Fleming}, {Johns-Krull}, {Phan-Bao},
  \& {Sherry}}]{2008ApJ...676.1307B}
{Berger}, E., {Basri}, G., {Gizis}, J.~E., {Giampapa}, M.~S., {Rutledge},
  R.~E., {Liebert}, J., {Mart{\'{\i}}n}, E., {Fleming}, T.~A., {Johns-Krull},
  C.~M., {Phan-Bao}, N., \& {Sherry}, W.~H. 2008, \apj, 676, 1307

\bibitem[{{Berger} {et~al.}(2009){Berger}, {Rutledge}, {Phan-Bao}, {Basri},
  {Giampapa}, {Gizis}, {Liebert}, {Mart{\'{\i}}n}, \&
  {Fleming}}]{2009ApJ...695..310B}
{Berger}, E., {Rutledge}, R.~E., {Phan-Bao}, N., {Basri}, G., {Giampapa},
  M.~S., {Gizis}, J.~E., {Liebert}, J., {Mart{\'{\i}}n}, E., \& {Fleming},
  T.~A. 2009, \apj, 695, 310

\bibitem[{{Berger} {et~al.}(2005){Berger}, {Rutledge}, {Reid}, {Bildsten},
  {Gizis}, {Liebert}, {Mart{\'{\i}}n}, {Basri}, {Jayawardhana}, {Brandeker},
  {Fleming}, {Johns-Krull}, {Giampapa}, {Hawley}, \&
  {Schmitt}}]{2005ApJ...627..960B}
{Berger}, E., {Rutledge}, R.~E., {Reid}, I.~N., {Bildsten}, L., {Gizis}, J.~E.,
  {Liebert}, J., {Mart{\'{\i}}n}, E., {Basri}, G., {Jayawardhana}, R.,
  {Brandeker}, A., {Fleming}, T.~A., {Johns-Krull}, C.~M., {Giampapa}, M.~S.,
  {Hawley}, S.~L., \& {Schmitt}, J.~H.~M.~M. 2005, \apj, 627, 960

\bibitem[{{Burgasser} {et~al.}(2000){Burgasser}, {Kirkpatrick}, {Reid},
  {Liebert}, {Gizis}, \& {Brown}}]{2000AJ....120..473B}
{Burgasser}, A.~J., {Kirkpatrick}, J.~D., {Reid}, I.~N., {Liebert}, J.,
  {Gizis}, J.~E., \& {Brown}, M.~E. 2000, \aj, 120, 473

\bibitem[{{Burgasser} \& {Putman}(2005)}]{2005ApJ...626..486B}
{Burgasser}, A.~J., \& {Putman}, M.~E. 2005, \apj, 626, 486

\bibitem[{{Burgasser} {et~al.}(2011){Burgasser}, {Sitarski}, {Gelino},
  {Logsdon}, \& {Perrin}}]{2011ApJ...739...49B}
{Burgasser}, A.~J., {Sitarski}, B.~N., {Gelino}, C.~R., {Logsdon}, S.~E., \&
  {Perrin}, M.~D. 2011, \apj, 739, 49

\bibitem[{{Ciliegi} {et~al.}(2003){Ciliegi}, {Zamorani}, {Hasinger}, {Lehmann},
  {Szokoly}, \& {Wilson}}]{2003A&A...398..901C}
{Ciliegi}, P., {Zamorani}, G., {Hasinger}, G., {Lehmann}, I., {Szokoly}, G., \&
  {Wilson}, G. 2003, \aap, 398, 901

\bibitem[{{Condon} {et~al.}(1998){Condon}, {Cotton}, {Greisen}, {Yin},
  {Perley}, {Taylor}, \& {Broderick}}]{1998AJ....115.1693C}
{Condon}, J.~J., {Cotton}, W.~D., {Greisen}, E.~W., {Yin}, Q.~F., {Perley},
  R.~A., {Taylor}, G.~B., \& {Broderick}, J.~J. 1998, \aj, 115, 1693

\bibitem[{{Dulk}(1985)}]{1985ARA&A..23..169D}
{Dulk}, G.~A. 1985, \araa, 23, 169

\bibitem[{{Fleming} {et~al.}(1995){Fleming}, {Schmitt}, \&
  {Giampapa}}]{1995ApJ...450..401F}
{Fleming}, T.~A., {Schmitt}, J.~H.~M.~M., \& {Giampapa}, M.~S. 1995, \apj, 450,
  401

\bibitem[{{Fuhrmeister} {et~al.}(2005){Fuhrmeister}, {Schmitt}, \&
  {Hauschildt}}]{2005A&A...439.1137F}
{Fuhrmeister}, B., {Schmitt}, J.~H.~M.~M., \& {Hauschildt}, P.~H. 2005, \aap,
  439, 1137

\bibitem[{{Gizis}(2002)}]{2002ApJ...575..484G}
{Gizis}, J.~E. 2002, \apj, 575, 484

\bibitem[{{Gizis} {et~al.}(2000){Gizis}, {Monet}, {Reid}, {Kirkpatrick},
  {Liebert}, \& {Williams}}]{2000AJ....120.1085G}
{Gizis}, J.~E., {Monet}, D.~G., {Reid}, I.~N., {Kirkpatrick}, J.~D., {Liebert},
  J., \& {Williams}, R.~J. 2000, \aj, 120, 1085

\bibitem[{{Greisen}(2003)}]{2003ASSL..285..109G}
{Greisen}, E.~W. 2003, Information Handling in Astronomy - Historical Vistas,
  285, 109

\bibitem[{{Guedel} \& {Benz}(1993)}]{1993ApJ...405L..63G}
{Guedel}, M., \& {Benz}, A.~O. 1993, \apjl, 405, L63

\bibitem[{{Hall}(2002)}]{2002ApJ...564L..89H}
{Hall}, P.~B. 2002, \apjl, 564, L89

\bibitem[{{Hallinan} {et~al.}(2008){Hallinan}, {Antonova}, {Doyle}, {Bourke},
  {Lane}, \& {Golden}}]{2008ApJ...684..644H}
{Hallinan}, G., {Antonova}, A., {Doyle}, J.~G., {Bourke}, S., {Lane}, C., \&
  {Golden}, A. 2008, \apj, 684, 644

\bibitem[{{Hallinan} {et~al.}(2007){Hallinan}, {Bourke}, {Lane}, {Antonova},
  {Zavala}, {Brisken}, {Boyle}, {Vrba}, {Doyle}, \&
  {Golden}}]{2007ApJ...663L..25H}
{Hallinan}, G., {Bourke}, S., {Lane}, C., {Antonova}, A., {Zavala}, R.~T.,
  {Brisken}, W.~F., {Boyle}, R.~P., {Vrba}, F.~J., {Doyle}, J.~G., \& {Golden},
  A. 2007, \apjl, 663, L25

\bibitem[{{James} {et~al.}(2000){James}, {Jardine}, {Jeffries}, {Randich},
  {Collier Cameron}, \& {Ferreira}}]{2000MNRAS.318.1217J}
{James}, D.~J., {Jardine}, M.~M., {Jeffries}, R.~D., {Randich}, S., {Collier
  Cameron}, A., \& {Ferreira}, M. 2000, \mnras, 318, 1217

\bibitem[{{Johns-Krull} \& {Valenti}(1996)}]{1996ApJ...459L..95J}
{Johns-Krull}, C.~M., \& {Valenti}, J.~A. 1996, \apjl, 459, L95

\bibitem[{{Kuznetsov} {et~al.}(2012){Kuznetsov}, {Doyle}, {Yu}, {Hallinan},
  {Antonova}, \& {Golden}}]{2012ApJ...746...99K}
{Kuznetsov}, A.~A., {Doyle}, J.~G., {Yu}, S., {Hallinan}, G., {Antonova}, A.,
  \& {Golden}, A. 2012, \apj, 746, 99

\bibitem[{{Liebert} {et~al.}(2003){Liebert}, {Kirkpatrick}, {Cruz}, {Reid},
  {Burgasser}, {Tinney}, \& {Gizis}}]{2003AJ....125..343L}
{Liebert}, J., {Kirkpatrick}, J.~D., {Cruz}, K.~L., {Reid}, I.~N., {Burgasser},
  A., {Tinney}, C.~G., \& {Gizis}, J.~E. 2003, \aj, 125, 343

\bibitem[{{Machado} {et~al.}(1980){Machado}, {Avrett}, {Vernazza}, \&
  {Noyes}}]{1980ApJ...242..336M}
{Machado}, M.~E., {Avrett}, E.~H., {Vernazza}, J.~E., \& {Noyes}, R.~W. 1980,
  \apj, 242, 336

\bibitem[{{McLean} {et~al.}(2011){McLean}, {Berger}, {Irwin}, {Forbrich}, \&
  {Reiners}}]{2011ApJ...741...27M}
{McLean}, M., {Berger}, E., {Irwin}, J., {Forbrich}, J., \& {Reiners}, A. 2011,
  \apj, 741, 27

\bibitem[{{McLean} {et~al.}(2012){McLean}, {Berger}, \&
  {Reiners}}]{2012ApJ...746...23M}
{McLean}, M., {Berger}, E., \& {Reiners}, A. 2012, \apj, 746, 23

\bibitem[{{Mohanty} {et~al.}(2002){Mohanty}, {Basri}, {Shu}, {Allard}, \&
  {Chabrier}}]{2002ApJ...571..469M}
{Mohanty}, S., {Basri}, G., {Shu}, F., {Allard}, F., \& {Chabrier}, G. 2002,
  \apj, 571, 469

\bibitem[{{Neupert}(1968)}]{1968ApJ...153L..59N}
{Neupert}, W.~M. 1968, \apjl, 153, L59

\bibitem[{{Osten} {et~al.}(2006){Osten}, {Hawley}, {Bastian}, \&
  {Reid}}]{2006ApJ...637..518O}
{Osten}, R.~A., {Hawley}, S.~L., {Bastian}, T.~S., \& {Reid}, I.~N. 2006, \apj,
  637, 518

\bibitem[{{Osten} {et~al.}(2009){Osten}, {Phan-Bao}, {Hawley}, {Reid}, \&
  {Ojha}}]{2009ApJ...700.1750O}
{Osten}, R.~A., {Phan-Bao}, N., {Hawley}, S.~L., {Reid}, I.~N., \& {Ojha}, R.
  2009, \apj, 700, 1750

\bibitem[{{Phan-Bao} {et~al.}(2007){Phan-Bao}, {Osten}, {Lim}, {Mart{\'{\i}}n},
  \& {Ho}}]{2007ApJ...658..553P}
{Phan-Bao}, N., {Osten}, R.~A., {Lim}, J., {Mart{\'{\i}}n}, E.~L., \& {Ho},
  P.~T.~P. 2007, \apj, 658, 553

\bibitem[{{Ravi} {et~al.}(2011){Ravi}, {Hallinan}, {Hobbs}, \&
  {Champion}}]{2011ApJ...735L...2R}
{Ravi}, V., {Hallinan}, G., {Hobbs}, G., \& {Champion}, D.~J. 2011, \apjl, 735,
  L2

\bibitem[{{Reiners} \& {Basri}(2010)}]{2010ApJ...710..924R}
{Reiners}, A., \& {Basri}, G. 2010, \apj, 710, 924

\bibitem[{{Reiners} \& {Christensen}(2010)}]{2010A&A...522A..13R}
{Reiners}, A., \& {Christensen}, U.~R. 2010, \aap, 522, A13+

\bibitem[{{Route} \& {Wolszczan}(2012)}]{2012ApJ...747L..22R}
{Route}, M., \& {Wolszczan}, A. 2012, \apjl, 747, L22

\bibitem[{{Schmidt} {et~al.}(2007){Schmidt}, {Cruz}, {Bongiorno}, {Liebert}, \&
  {Reid}}]{2007AJ....133.2258S}
{Schmidt}, S.~J., {Cruz}, K.~L., {Bongiorno}, B.~J., {Liebert}, J., \& {Reid},
  I.~N. 2007, \aj, 133, 2258

\bibitem[{{Schrijver}(2009)}]{2009ApJ...699L.148S}
{Schrijver}, C.~J. 2009, \apjl, 699, L148

\bibitem[{{Stelzer} {et~al.}(2006){Stelzer}, {Micela}, {Flaccomio},
  {Neuh{\"a}user}, \& {Jayawardhana}}]{2006A&A...448..293S}
{Stelzer}, B., {Micela}, G., {Flaccomio}, E., {Neuh{\"a}user}, R., \&
  {Jayawardhana}, R. 2006, \aap, 448, 293

\bibitem[{{Wilson} {et~al.}(2011){Wilson}, {Ferris}, {Axtens}, {Brown},
  {Davis}, {Hampson}, {Leach}, {Roberts}, {Saunders}, {Koribalski}, {Caswell},
  {Lenc}, {Stevens}, {Voronkov}, {Wieringa}, {Brooks}, {Edwards}, {Ekers},
  {Emonts}, {Hindson}, {Johnston}, {Maddison}, {Mahony}, {Malu}, {Massardi},
  {Mao}, {McConnell}, {Norris}, {Schnitzeler}, {Subrahmanyan}, {Urquhart},
  {Thompson}, \& {Wark}}]{2011MNRAS.416..832W}
{Wilson}, W.~E., {Ferris}, R.~H., {Axtens}, P., {Brown}, A., {Davis}, E.,
  {Hampson}, G., {Leach}, M., {Roberts}, P., {Saunders}, S., {Koribalski},
  B.~S., {Caswell}, J.~L., {Lenc}, E., {Stevens}, J., {Voronkov}, M.~A.,
  {Wieringa}, M.~H., {Brooks}, K., {Edwards}, P.~G., {Ekers}, R.~D., {Emonts},
  B., {Hindson}, L., {Johnston}, S., {Maddison}, S.~T., {Mahony}, E.~K.,
  {Malu}, S.~S., {Massardi}, M., {Mao}, M.~Y., {McConnell}, D., {Norris},
  R.~P., {Schnitzeler}, D., {Subrahmanyan}, R., {Urquhart}, J.~S., {Thompson},
  M.~A., \& {Wark}, R.~M. 2011, \mnras, 416, 832

\bibitem[{{Wu} \& {Lee}(1979)}]{1979ApJ...230..621W}
{Wu}, C.~S., \& {Lee}, L.~C. 1979, \apj, 230, 621

\bibitem[{{Zarka} {et~al.}(2001){Zarka}, {Queinnec}, \&
  {Crary}}]{2001P&SS...49.1137Z}
{Zarka}, P., {Queinnec}, J., \& {Crary}, F.~J. 2001, \planss, 49, 1137

\end{thebibliography}

\end{document}